# INFORMATION TOKEN DRIVEN MACHINE LEARNING FOR ELECTRONIC MARKETS: PERFORMANCE EFFECTS IN BEHAVIORAL FINANCIAL BIG DATA ANALYTICS

**Jim Samuel**
jim@aiknowledgecenter.com

**ABSTRACT**

Conjunct with the universal acceleration in information growth, financial services have been immersed in an evolution of information dynamics. It is not just the dramatic increase in volumes of data, but the speed, the complexity and the unpredictability of 'big-data' phenomena that have compounded the challenges faced by researchers and practitioners in financial services. Math, statistics and technology have been leveraged creatively to create analytical solutions. Given the many unique characteristics of financial bid data (FBD) it is necessary to gain insights into strategies and models that can be used to create FBD specific solutions. Behavioral finance data, a subset of FBD, is seeing exponential growth and this presents an unprecedented opportunity to study behavioral finance employing big data analytics methodologies. The present study maps machine learning (ML) techniques and behavioral finance categories to explore the potential for using ML techniques to address behavioral aspects in FBD. The ontological feasibility of such an approach is presented and the primary purpose of this study is propositioned: ML based behavioral models can effectively estimate performance in FBD. A simple machine learning algorithm is successfully employed to study behavioral performance in an artificial stock market to validate the propositions.

**Keywords**: Information, Big Data, Electronic Markets, Analytics, Behavior

jim@aiknowledgecenter.com

# INTRODUCTION

Exploring big data about big data, in their editorial for Information Systems Research journal, Agarwal and Dhar (2014) reported their August-2014 Google search findings for the phrases "Big data," "Analytics," and "Data science" which generated 822 million, 154 million, and 461 million results, respectively. A similar search in February of 2017 for the present study has yielded 832 million results for the term "Analytics" and 93.4 million results for the phrase "Financial Analytics"! Variance in number of search results search results trends have been used for successfully predicting election results, stock performance, health care trends and customer sentiment. These metrics, though vague explorative indicators, are indicators nevertheless of the relatively growing importance and prominence of analytics and very specifically 'financial analytics', which measures over 11% of the search results for 'analytics'. Applying a similar search for "artificial intelligence" (AI) and "machine learning" (ML) in February of 2017 has yielded 89.6 million and 31.5 million results respectively. More enlightening is the "search trend" which charts the number of searches for a particular term or phrase. Running a five year search trend analysis for the terms "artificial intelligence" and "machine learning" showed limited increase in searches for AI but an almost exponential looking upward growth curve for searches in ML. Business and societal challenges, including 'financial markets stability' and 'complex challenges' (Ketter et. al., 2016), need to be tackled by going beyond traditional methods and tools into big data analytics methodologies and resources. Undoubtedly, given the domain impact implications along with rising global interest as indicated above, 'financial analytics', 'big data' and 'machine learning' are critically relevant phenomena which are in need of significant research attention so that we can gain insights into optimal management and value creation.

The paper is organized into four sections after the introduction: the growing relevance of behavioral financial big data, development of propositions, ML application to identify financial behavior and closes with the discussion and conclusion section. This is followed by sections on behavioral finance, machine learning and the intersection of behavioral finance and machine learning which is the focus of the present study. Using an ontological framework, the primary proposition is then developed and presented. An explorative study is used to demonstrate the viability of the primary proposition posited in the study, followed by concluding notes.

# FINANCIAL BIG DATA

An NVP (2016) survey of Fortune 1000 companies shows that big data analytics is now an important and established part of business strategy: Less than 6% of responding companies, in their 2014 survey, aimed to invest more than $50MM into big data projects. That number has grown from under 6% to over 26% of companies who intend to invest more than $ 50 million into big data analytics by 2017. Leading big data and analytics services companies such as Splunk, Talend, Hortonworks & New Relic have reported around 40% growth for 2016.The financial services sector has been one of the leading consumers of big data services. In their study focused on FBD analytics, IBM (2012) reported significant efforts by banks and companies to scale up their analytics initiatives and the acquisition of more data through internal and external information sources. Banks and large financial services firms have primarily used big data for exploratory analytics and insights in four areas: risk mitigation, expansion of traditional portfolio management, client and customer intelligence, and regulatory goals. Smaller financial services firms are in a position to create greater value using predictive big data analytics for innovative purposes to drive revenue and profitability.

Of the many dimensions of FBD, some of which have mature big data analytics models, the behavioral finance dimension of FBD has not been addressed with sufficient clarity by researchers, nor by practitioners, in context of big data phenomena. Behavioral finance gained attention in the 1990's (Barberis & Thaler, 2003) and is now treated as an established research domain. Behavioral finance is defined at the intersection of finance and psychology as the science of the *" ... application of psychology to finance, with a focus on individual-level cognitive biases*" (Hirshleifer, 2015). Taking an information systems perspective, the Editor-in-Chief of MISQ emphasized the critical role of behavioral analytics across domains stating "*Precise capture of individual behavior and surrounding events also allows for spotting population trends and the impact of events..*" (Goes, 2014). Behavioral effects in financial markets are driven by cognitive biases (Barberis & Thaler, 2003), emotions and feelings(O'Creevy, *et al.*, 2011; Cipriani, *et al.*) and personalized purposes or subjective expectations (Schwartz, *et al.*, 2010; Sahi, *et al.*, 2013). With the advent of internet-of-things (IoT) and its promise of rich voluminous continuous real-time high quality consumer behavior data, FBD is set to explode in quantity, acceleration of data acquisition, types of data and the essential distinctiveness of characteristics of data – aligned with the four V's of big data – Volume, Velocity, Variety and Veracity (Chen, *et al.*, 2012). It becomes clear that behavioral finance, across its spectrum, will be driven and shaped by big data phenomena: root level investor and trader behavior will be impacted by big data, corporations will be overwhelmed with the expanded FBD that becomes available, governments will seek to leverage behavioral FBD for regulatory purposes and researchers will have tremendous opportunities to provide thought leadership using innovative FBD analytics based strategies. This perspective is supported by Hirshleifer (2015) who emphasizes the need for a shift in focus for future studies in behavioral finance, a latent acknowledgement of the influence of big data in society, by calling for behavioral finance to go expand itself into " *... social finance, which studies the structure of social interactions, how financial ideas spread and evolve, and how social processes affect financial outcomes*". Thus research, big data phenomena and industry trends for financial services organizations indicate that behavioral FBD (BFBD) analytics will be a critical source of competitive advantage.

Statistical learning is not a newly developed research or application domain - it started in the 1960's, remained mostly theoretical till the 1990's and then gained momentum with the development of new algorithms which made practical application viable (Vapnik, 1999). However, the full potential of statistical learning applications only began to be realized with the advent of three forces: cross-domain development of algorithmic models, significant scaling up of processing capabilities, and big data phenomena. The unique confluence of these three powerful forces over the past decade, at a 'kairos' moment in human-technological progress, have led to an extraordinary acceleration in the science and application of statistical learning which has morphed with vast reach under the broad title of "Machine Learning" (ML). Machine learning is itself considered a subset on artificial intelligence (AI), but at the point of the writing of this study, ML is the science of relatively popular interest for big data analytics and organizational application. Specialized sub-domains such as "deep learning" have also been gaining traction. Machine learning has been defined in various ways, reflecting various application perspectives – Murphy (2012) defines ML by its purpose stating "*The goal of machine learning is to develop methods that can automatically detect patterns in data, and then to use the uncovered patterns to predict future data or other outcomes of interest*". Alpaydin (2014) defines ML as " … *programming computers to optimize a performance criterion using example data or past experience*". ML can be classified into three primary categories: supervised learning – learns to predicts values, unsupervised learning – learns to find similar instances, and reinforcement learning – learn to improve based on sequence outcomes. Other categories such as active learning, hybrid learning, deep learning and semi-supervised learning are also significant. While domain specific narratives and definitions abound, ML can be summarized as being the science of using mathematical and statistical methods, technology tools, domain knowledge and information management techniques to identify patterns, positions, relationships and designs in information artifacts ranging along the continuum from simple tabular data to complex unstructured data.

## PROPOSITION DEVELOPMENT

The present study posits that BFBD analytics can leverage ML techniques to create insights on user behaviors and thus create meaningful avenues for high quality customer intelligence and trading performance. It must be noted that the traditional monolithic definition of "information", which embodied the assumption of uniform objective interpretation and rational utility maximization led many behavioral finance researchers to discount information as being of little use in understanding behavioral phenomena (Barberis and Thaler, 2003). However subsequent research has shown that information influences financial decisions and behavior (Garcia, 2013), albeit the notion of "information" must expanded (Grover, 2006) beyond uniform objectivity and rationality for all financial decisions. Psychologists have distinguished between information, knowledge, belief and behavior, but taken conflicting positions of role of knowledge and information in human decision making (Ajzen, *et al.*, 2011). Recent advances in information theory and domain specific perspectives of information present information as an evolutionary multi-dimensional construct with significant behavioral implications (Floridi, 2011). The complexity of the "information" construct has been discussed in information systems research and an integrative taxonomy, explorative in nature, has been proposed (McKinney, 2010). While epistemological development has progressed, we do not have any objective ontological mapping of "information" as a construct and this makes "information" a very fuzzy construct to deal with theoretically. The present study does not attempt to define or provide any ontological analysis for "information" but it simply purposes to highlight the associated complexity and move on to define two terms introduced in the present study "Information Virtue" and "Information Token" which are relatively tangible constructs with significant relevance for FBD analytics. "Information Token" is used akin to the "token" concept introduced by McKinney (McKinney, *et al.*, 2010), who define information from a "token" perspective as being "… *an undifferentiated commodity of data bits that are processed, not a particular relation among the bits*". The present study defines "Information Token" ($\tau$) as a tangible, storable and transferrable expression of data in an objective manner from an entity-neutral perspective. "Information Virtue" is defined as the ultimate meaning of specific information from a source-of-information perspective. This means that Information Virtue ($\alpha$) can be treated as a distinct ultimate meaning (source perspective) even though different persons or entities may find different meanings (destination perspective) or process it based on subjective models (process perspective) and reach varying conclusions. Thus a source perspective Information Virtue ($\alpha$) can be fairly expected, based on theoretical perspectives presented above, to lead to similar (subject to variance in properties of Information Token) or differentiated process or destination perspectives. Information influences beliefs and behavior (Ajzen and Fishbein, 1980) – the decisions stimulated by the Information Token expressing the Information Virtue will affect beliefs and lead to information driven performance. Here, we address this individual level (or homogenous entity, defined by common Information Token ($\tau$)) level behavior as "Informational Performance" ($\rho$). It is possible for Information Virtue ($\alpha$) to vary ($\alpha i \rightarrow \alpha k$) and then it is not surprising that such a variance in α would lead to changes in Informational Performance (ρ) and it would contain no implication for behavioral finance. However if Information Virtue (α) is fixed and Information Token (τ) is varied, then any corresponding change in as Informational Performance (ρ) would have significant implication for research behavioral finance and for financial service practitioners. This conceptual position is tabulated below and then discussed.

**Table 1.** Information Token-Performance Table.

| Virtue | Token | Performance |
|---|---|---|
| αi | τ 1 | ρ 1 |
| αi | τ 2 | ρ 2 |
| αi | τ 3 | ρ 3 |
| αi… | τ … | ρ … |
| αi | τ n | ρ n |

Traditional perspectives posited that for any specific Information Virtue (αi), based on uniform rational expectations, there will be no variance in Informational Performance (ρn) irrespective of variance of Information Token (τn). This implies that, for a given (αi), change in Information Token "Δ (τn)" does not affect (ρn). In contrast, the present study posits that for a given (αi), (ρn) will vary with Δ (τn), assuming Δ (τn) represents sufficient variance in expression of the same (αi). Thus (ρn) is expected to be a function of (τn) for a specific (αi), given by the equation:

$$f(g(\tau n)) \equiv E(\rho n \mid \alpha i) \tag{1}$$

Where, E is the expectation of ρn for a specific αi, given by the function of *g* (τn), where *g* (τn) represents the sufficient variance function of τn. This prepares the context for the addressing the classification of behavior given by the functions of (τn) based on Informational Performance (ρn). It has been amply demonstrated by research in psychology (Ajzen, *et al.*, 2011), information systems (Delone and McLean, 2003), management (Bentzen, *et al.*, 2011) and finance (De Bondt, *et al.*, 2013) that information affects behavior, positioning Information Token (τn) as a direct antecedent to behavior, which is often a directly unobservable latent construct. The decision variable is more tangible is a direct outcome of behavior and thus Informational Performance (ρ) serves as a proxy for understanding Information Token (τn) driven behavior. Thus the above theoretical discussion leads us to the primary proposition of our paper:

> Primary Proposition (Pa): "*For a specific Information Virtue (αi), sufficient variance in Information Token (τn) will lead to differentiated behavior, given by levels of Informational Performance (ρn)*"

Machine Learning has been used to study and improve complex decision environments (Meyer, *et al.*, 2014), emotional state classification in neurological studies (Wang, *et al.*, 2014) and various types of text mining based market predictions (Nassirtoussi, *et al.*, 2014). Google search and Twitter tweet trends, along with a host of social media generated big data has also been used to understand and predict, at a high level, stock price and market movements. However, there has been a dearth of application of machine learning techniques to study individual behavioral responses to information artifacts. Therefore, in the present study, we leverage ML capabilities to posit that ML classification techniques can be used to categorize individual financial performance reflecting behavioral effects in response to sufficient variance to Information Tokens (τn). This leads to the secondary proposition of this study:

> Secondary Proposition (Pb): "*For a specific Information Virtue (αi), behavior driven by sufficient variance in Information Tokens (τn) can be classified by machine learning classification algorithms applied to Informational Performance (ρn).*

The secondary proposition (Pb) is explored in the present study by using performance data from an artificial stock market. However, the concluding proposition is developed and stated on an *a priori* basis, using inductive logic from the above discussion, combined with the logic of non-parametric analytical possibilities that are now available through machine learning and other advanced mathematical techniques, which provide useful results even with unknown probabilistic distributions of data. Machine learning techniques have been leveraged to detect patterns (Fischer and Igel, 2014) with latent variables across multiple levels using Deep Learning methods such as Deep Belief Networks (DBN) using restricted Boltzmann machines (RBMs). Machine learning and data mining techniques using Bayesian Belief Networks (BNN), and fuzzy logic and genetic algorithms have been used extensively even in the absence of complete information or sufficient prior probabilities to solve a variety of financial services challenges, especially fraud detection (Sharma and Panigrahi, 2012) related patterns and behavior. Neural networks, decision trees, simulation and optimization techniques have been used extensively to develop more efficient solutions to financial markets challenges. The third and concluding proposition is built on the knowledge of these capabilities, without which it would be suspect as conjecture, that ML techniques can be used to model behavior affecting financial performance just as it has already been used to classify behavior associated with financial fraud. The obvious implication, developing upon Pa and Pb, is that we use advanced algorithmic, intelligent network and fuzzy logic methods to to classify, interpret and predict expected financial performance using BFBD. This brings us to the final proposition of this study:

> Concluding Proposition (Pf): "*Machine learning behavioral models, based on the function of sufficient variance in Information Tokens (f(g($\tau_n$)), can effectively estimate expected performance (E ( ρn | αi )) using behavioral financial big data*".

The mathematical concept underlying the final proposition can be expressed in the context of a collection of matrices {[F...]}, not necessarily a sum, of application of relevant machine learning techniques (F) to a sufficiently varied Information Token (, which is given by the equation:

$$\{[F*(f(g(\tau_n)))]\} \equiv E (\rho n | \alpha i) \qquad (2)$$

In a simple scenario such as a controlled artificial electronic market, where an instance of application of a machine learning technique (Fn) to sufficiently varied Information Token *(τn)*, suffices the necessary analytics condition, then equation E2 can be simplified thus:

$$Fn*(f(g(\tau_n) E (\rho n | \alpha i) \qquad (3)$$

Based on the above, the present study uses data from an artificial stock market to study how individual level Informational Performance (ρ) data can be trained to classify behavioral responses given by the function of Information Token (τ). This leads to the next section where a parsimonious application of a simple ML method "k-Nearest Neighbors (KNN) algorithm" classifies behavior based on distinct information artifacts representing Information Tokens (τn). The mathematical concept used for framing the analysis can be expressed as an application of equation E1, where a specific function (KNN) is used on individual trading Informational Performance (ρ(trading)) to classify a set of seven Information Token {(τn(1:7))} conditions, given by the equation:

$$f_{knn}\ (\rho n(trading)\ |\ \alpha i) \equiv E\ (g(\tau n)) \tag{4}$$

Information Virtue (αi) is held constant (source perspective) based on ensuring that the effective meaning is unchanged (such as "stock price for company 'X' will increase today"). Information Tokens (τn) are varied using seven representational artifacts.

**MACHINE LEARNING APPLICATION TO CLASSIFY FINANCIAL BEHAVIOR**

A simple ML method k-Nearest Neighbors (KNN) algorithm is used to classify artificial stock market trading Informational Performance. Data is drawn from a series of equity trading conducted with student subjects from a large Northeast US University. Subjects were undergraduate students with limited prior exposure to electronic stock trading. Equity trading was simulated using the TraderEx (2016) platform which has been widely for simulations, teaching and training electronic market dynamics by graduate and undergraduate faculty and equity market professionals. Economically motivated subjects were introduced to the Information Token conditions and tasked to trade with an intent to maximize end-of-day trading profits. Though the present nonparametric KNN algorithm based ML analytics do not mandate the rigor of a parametric approach, for the sake of internal validity, the sufficient distinctiveness of six Information Tokens were verified using manipulation checks using scales adapted from Dennis and Kinney (1998). This satisfies the necessary condition to test the data for behavioral effects. The Information Tokens were mutually distinguished using high and low levels of cognitive information based on deterministic, probabilistic and quantity of information. Subjects were subject to only one Information Token condition to avoid learning effects. The last Information Token *(τ7)* was a no information guidance condition (α0), used as a base to observe information effects with greater clarity.

**Table 2.** Data Organization For ML Analytics.

| Token | Performance | ρ Count | Train* | Test* |
|---|---|---|---|---|
| T1 | ρ 1 | 30 | 23 | 7 |
| T2 | ρ 2 | 35 | 26 | 9 |
| T3 | ρ 3 | 31 | 24 | 7 |
| T4 | ρ 4 | 30 | 20 | 10 |
| T5 | ρ 5 | 30 | 21 | 9 |
| T6 | ρ 6 | 34 | 26 | 8 |
| T7 | ρ 7 | 33 | 19 | 14 |
| Total | | 223 | 159 | 64 |

* Train and Test items, and therefore count, were set to be randomly selected by R, based on an approximately 7:3 ratio overall for test:train, and is tabulated here as output

For the given Information Virtue (αi), with the exception of the deliberate (α0) condition, which indicated a stock price increase between 2% and 5% in all the Information Token conditions:

$$g(\tau 1) \neq g(\tau 2) \neq g(\tau 3) \neq g(\tau 4) \neq g(\tau 5) \neq g(\tau 6) \neq g(\tau 7) \quad (5)$$

The analysis that follows tests if, given a specific Information Virtue (αi), behavior driven by sufficient variance in Information Tokens (τn) can be classified by the KNN algorithm applied to Informational Performance (ρn)?. A total of 223 Informational Performance (ρ) measures were recorded from these Information Tokens, given by net profits for the trading session by each subject. The data are divided into training dataset and testing dataset with a training to testing ratio of 0.7:0.3. R software is used to run the KNN algorithm using appropriate R packages and the raw output is available in the appendix. The financial trading behavioral effect, Informational Performance (ρn) is summarized by the equation:

$$(\rho 1) \neq (\rho 2) \neq (\rho 3) \neq (\rho 4) \neq (\rho 5) \neq (\rho 6) \neq (\rho 7) \quad (6)$$

The KNN algorithm is expected to 'train' the model using the training dataset and apply it to the test dataset by classifying Informational Performance (ρn) measures to corresponding Information Token (τn) conditions. This set of classifications is given by the equation

$$f_{knn}(\rho(1:7) \mid \alpha i) \rightarrow \{ C1(g(\tau 1)), C2(g(\tau 2)), C3(g(\tau 3)), C4(g(\tau 4)), C5(g(\tau 5)), C6(g(\tau 6)), C7(g(\tau 7)) \} \quad (7)$$

Where each Cn represents a unique classification of behavior given by $g(\tau n)$. The usefulness and the accuracy of the KNN algorithm increases with the increase in the size of the training dataset and is therefore very suitable for BFDB analytics, subject to it being mathematically adapted for speedy large scale implementation. Using 159 training items, applying the KNN algorithm on 64 test items demonstrated that the algorithm was able to correctly classify the behavior by associating the Information Tokens to corresponding Informational Performance in 57 of the 64 cases, leading to a success rate of 89%.

Table 3. KNN Algorithm Classification Output Comparison.

| KNN Algorithm Classification Output Comparison – Actual : Classification * | | | | | | | | |
|---|---|---|---|---|---|---|---|---|
| | Classification | | | | | | | |
| Actual | T1 cl | T2 cl | T3 cl | T4 cl | T5 cl | T6 cl | T7 cl | Row Total |
| T1 | 7 | 0 | 0 | 0 | 0 | 0 | 0 | 7 |
| T2 | 0 | 9 | 0 | 0 | 0 | 0 | 0 | 9 |
| T3 | 0 | 0 | 6 | 0 | 0 | 0 | 1 | 7 |
| T4 | 0 | 0 | 0 | 9 | 0 | 1 | 0 | 10 |
| T5 | 0 | 0 | 0 | 0 | 9 | 0 | 0 | 9 |
| T6 | 0 | 1 | 0 | 0 | 0 | 7 | 0 | 8 |
| T7 | 4 | 0 | 0 | 0 | 0 | 0 | 10 | 14 |
| Column Total | 11 | 10 | 6 | 9 | 9 | 8 | 11 | 64 |

*\* Raw R output is provided in the Appendix*

Also, four of the seven erroneous items were are artifact of Information Token 7 which was the no-information token condition (T7) and hence looking at T1:T6, we see a stronger success rate of 94%, with 47 successful and 3 missed classifications.

**Table 4.** ML Analytics Summary.

| | Summary of KNN Classification Success | | |
|---|---|---|---|
| | Success | Missed | Success% |
| T1:T7 | 57 | 7 | 89% |
| T1:T6 | 47 | 3 | 94% |

## DISCUSSION AND CONCLUSION

The present study has attempted to highlight three key propositions at the intersection of FBD, behavioral finance, application of analytics and the use of machine learning to classify behavior. In doing so, it has combined perspectives from information systems, behavioral finance, electronic markets and big data analytics. A new and unique perspective for identifying behavioral classifications based on Information Virtue, Information Tokens and Informational Performance has been developed and validated using a ML KNN algorithm methodology. The propositions of the present study are mainly relevant to electronic financial markets, in studying behavior in the context of information stimuli. However, the propositions could be generalized and the underlying principles could be used for understanding big data analytics in other domains.

The study has certain limitations. The primary concern, as it is with most experimental studies and studies using simulated data, is that of external validity – the data obtained from the artificial stock market used student subjects with limited prior understanding of electronic equity markets and it can be claimed that such subjects are more vulnerable to Information Token manipulations than professional traders and investors. However, given the benefit of a controlled artificial electronic market setting, the effects of the Information Artifacts could be clearly delineated. Another limitation is the use of a single algorithm in a single instance – it has not been the purpose of the present study to provide an exhaustive and elaborate analysis of ML techniques, but to highlight the possibilities associated with using ML methodologies (Meyer, *et al.*, 2014) in addressing financial behavior.

The machine learning approach and the Information Virtue – Information Token – Informational Performance perspective presented here can be can be used, in addition to applications in BFBD, to address a wide range of behavioral challenges in electronic markets. Many quasi-financial electronic marketplaces are highly behavioral, such as markets for people-to-people lending with a strong trust-behavior influence (Greiner and Wang, 2010), where the analytical approach described in this study can be effectively applied to reduce risk and increase profitability. Furthermore, in complex informational environments surrounding electronic market places, machine learning can be effectively applied to identify patterns not discoverable by traditional business intelligence methods. For example, early studies in micro-commoditization pointed to information growth and increased complexity of information in electronic market environments (Gopal, *et al.*, 2003). Similarly, online marketplaces have seen a steady flow of creative phenomena such as social shopping communities and clickstream data (Olbrich and Holsing, 2011). Traditional business intelligence would not have the tools or the strategies required to address complex electronic market dynamics resulting from phenomena such as micro-commoditization. As expressed in the present study, ML techniques and emerging information management models will be more effective in tackling such challenges and the findings of the present study are expected to serve as a contribution in this direction.

In conclusion, the present study presents a novel approach to behavioral analytics using information stimuli (Virtue), information artifacts (Tokens) driven behavior and corresponding performance measures. Three unique propositions were developed using inductive logic demonstrating how ML behavioral models, based on the function of sufficient variance in Information Tokens can effectively estimate expected performance – simply stated: ML behavioral models can be trained to effectively estimate performance in FBD. In a parsimonious validation, an application of the KNN algorithm demonstrated a fair measure of success in classifying information tokens. To the best of our knowledge, there is no other study that has addressed the issues pertaining to behavioral analytics using the information framework presented in this paper. Additional research will be required to further explore the propositions presented in this study, using empirical data, and particularly in expanding the depth and scope and algorithmic application in the context of the Information Virtue – Information Token – Informational Performance framework.

**APPENDIX**

R raw output of KNN algorithm, compared using library(gmodels)

CrossTable(x = gg1.testLabels, y = gg1p, prop.chisq=FALSE)

```
   Cell Contents
|-------------------------|
|                       N |
|           N / Row Total |
|           N / Col Total |
|         N / Table Total |
|-------------------------|
```

Total Observations in Table:  64

| gg1.testLabels | T1 | T2 | T3 | T4 | T5 | T6 | T7 | Row Total |
|---|---|---|---|---|---|---|---|---|
| T1 | 7 | 0 | 0 | 0 | 0 | 0 | 0 | 7 |
|    | 1.000 | 0.000 | 0.000 | 0.000 | 0.000 | 0.000 | 0.000 | 0.109 |
|    | 0.636 | 0.000 | 0.000 | 0.000 | 0.000 | 0.000 | 0.000 |  |
|    | 0.109 | 0.000 | 0.000 | 0.000 | 0.000 | 0.000 | 0.000 |  |
| T2 | 0 | 9 | 0 | 0 | 0 | 0 | 0 | 9 |
|    | 0.000 | 1.000 | 0.000 | 0.000 | 0.000 | 0.000 | 0.000 | 0.141 |
|    | 0.000 | 0.900 | 0.000 | 0.000 | 0.000 | 0.000 | 0.000 |  |
|    | 0.000 | 0.141 | 0.000 | 0.000 | 0.000 | 0.000 | 0.000 |  |
| T3 | 0 | 0 | 6 | 0 | 0 | 0 | 1 | 7 |
|    | 0.000 | 0.000 | 0.857 | 0.000 | 0.000 | 0.000 | 0.143 | 0.109 |
|    | 0.000 | 0.000 | 1.000 | 0.000 | 0.000 | 0.000 | 0.091 |  |
|    | 0.000 | 0.000 | 0.094 | 0.000 | 0.000 | 0.000 | 0.016 |  |
| T4           |   0   |   0   |   0   |   9   |   0   |   1   |   0   |  10 |
|              | 0.000 | 0.000 | 0.000 | 0.900 | 0.000 | 0.100 | 0.000 | 0.156 |
|              | 0.000 | 0.000 | 0.000 | 1.000 | 0.000 | 0.125 | 0.000 |     |
|              | 0.000 | 0.000 | 0.000 | 0.141 | 0.000 | 0.016 | 0.000 |     |
| T5           |   0   |   0   |   0   |   0   |   9   |   0   |   0   |   9 |
|              | 0.000 | 0.000 | 0.000 | 0.000 | 1.000 | 0.000 | 0.000 | 0.141 |
|              | 0.000 | 0.000 | 0.000 | 0.000 | 1.000 | 0.000 | 0.000 |     |
|              | 0.000 | 0.000 | 0.000 | 0.000 | 0.141 | 0.000 | 0.000 |     |
| T6           |   0   |   1   |   0   |   0   |   0   |   7   |   0   |   8 |
|              | 0.000 | 0.125 | 0.000 | 0.000 | 0.000 | 0.875 | 0.000 | 0.125 |
|              | 0.000 | 0.100 | 0.000 | 0.000 | 0.000 | 0.875 | 0.000 |     |
|              | 0.000 | 0.016 | 0.000 | 0.000 | 0.000 | 0.109 | 0.000 |     |
| T7           |   4   |   0   |   0   |   0   |   0   |   0   |  10   |  14 |
|              | 0.286 | 0.000 | 0.000 | 0.000 | 0.000 | 0.000 | 0.714 | 0.219 |
|              | 0.364 | 0.000 | 0.000 | 0.000 | 0.000 | 0.000 | 0.909 |     |
|              | 0.062 | 0.000 | 0.000 | 0.000 | 0.000 | 0.000 | 0.156 |     |
| Column Total |  11   |  10   |   6   |   9   |   9   |   8   |  11   |  64 |
|              | 0.172 | 0.156 | 0.094 | 0.141 | 0.141 | 0.125 | 0.172 |     |